\documentclass[11pt]{article}

\usepackage{epsfig,subfigure}
\usepackage{amssymb}
\usepackage[fleqn]{amsmath}
\usepackage{color}
\newcommand{\define}{\stackrel{\triangle}{=}}
\def\QED{\mbox{\rule[0pt]{1.5ex}{1.5ex}}}
\def\proof{\noindent\hspace{2em}{\it Proof: }}

\usepackage{graphicx}

\usepackage{amssymb}
\usepackage{amsmath,amsfonts,amssymb}
\usepackage{verbatim}
\usepackage{stfloats}
\usepackage[bookmarks=false]{}

\newtheorem{theorem}{Theorem}

\newtheorem{lemma}{Lemma}

\newcommand\blfootnote[1]{%
  \begingroup
  \renewcommand\thefootnote{}\footnote{#1}%
  \addtocounter{footnote}{-1}%
  \endgroup
}

\begin{document}
\date{}
\title{
The Capacity of Symmetric \\
 Private Information Retrieval
}
\author{\normalsize Hua Sun, Syed A. Jafar \\
}

\maketitle

\blfootnote{This paper was presented in part at NETCOD 2016. Hua Sun (email: huas2@uci.edu) 
and Syed A. Jafar (email: syed@uci.edu) are with the Center of Pervasive Communications and Computing (CPCC) in the Department of Electrical Engineering and Computer Science (EECS) at the University of California Irvine. }

\begin{abstract}
Private information retrieval (PIR) is the problem of retrieving as efficiently as possible, one out of $K$ messages from $N$ non-communicating replicated databases (each holds all $K$ messages) while keeping the identity of the desired message index a secret from each individual database. Symmetric PIR (SPIR) is a generalization of PIR to include the requirement that beyond the desired message, the user learns nothing about the other $K-1$ messages. The information theoretic capacity of SPIR (equivalently, the reciprocal of minimum download cost) is the maximum number of bits of desired information that can be privately retrieved per bit of downloaded information. We show that the capacity of SPIR is $1-1/N$ regardless of the number of messages $K$, if the databases have access to common randomness (not available to the user) that is independent of the messages, in the amount that is at least $1/(N-1)$ bits per desired message bit, and zero otherwise. 
Extensions to the capacity region of SPIR and the capacity of finite length SPIR are  provided.
\end{abstract}

\newpage
\allowdisplaybreaks
\section{Introduction}

The private information retrieval (PIR) problem \cite{PIRfirst, PIRfirstjournal} seeks the most efficient way for a user to retrieve a desired message from a set of distributed databases, each of which stores all the messages, without revealing any information about which message  is being retrieved to any individual database. This seemingly impossible mission has a trivial (expensive) solution, i.e., the user can request all the messages to hide his interest. The goal of the PIR problem is to find the most efficient solution. The capacity of PIR is defined as the maximum number of bits of desired message that can be privately downloaded per bit of downloaded information. In our recent work \cite{Sun_Jafar_PIR}, the capacity of PIR  with $K$ messages and $N$ databases was shown to be $C_{\footnotesize \mbox{PIR}} = (1 + 1/N + \cdots + 1/N^{K-1})^{-1}$.

The original formulation of PIR only considers the privacy of the user. The privacy of the undesired messages is ignored. However, it is often desirable to restrict the user to retrieve nothing beyond his chosen message. This new constraint is called database privacy, and with this constraint, the problem is called symmetric\footnote{Symmetry means that the privacy of both the user and the database is considered.} PIR (SPIR) \cite{SymPIR}. Symmetric PIR  is especially challenging because the databases must  individually learn nothing about the identity of the desired message, but must still collectively allow the user to retrieve his desired message in such a way that the user learns nothing about any other message besides his desired message. For example,  the trivial solution of downloading everything, is no longer acceptable.  The main contribution of this work is the characterization of the capacity of SPIR, i.e., the maximum number of bits of desired message that can be privately retrieved by a user per bit of downloaded information, without leaking any information about undesired messages to the user. For $K$ messages and $N$ databases, we show that the capacity is $1 - 1/N$. Extensions of the main result, from capacity to capacity region and from infinite message length to arbitrary message length, are also provided.

Besides its direct applications, PIR is especially significant as a fundamental problem that lies at the intersection of several open problems in cryptography \cite{William, Yekhanin}, coding theory \cite{LDC, YekhaninPhd, Batch} and complexity theory \cite{Ishai_Kushilevitz}. SPIR inherits many of these connections from PIR. For example, SPIR is essentially a (distributed) form of oblivious transfer \cite{Rabin, Even_OT}, where the typical objective is  that the transmitter(s) should not know which message is received by the receiver and the receiver should obtain nothing more than the desired message. Oblivious transfer is an important building block (primitive) in cryptography, whose feasibility leads to many other cryptographic protocols \cite{Killian, Ishai_Prabhakaran_Sahai}. Fundamental limits on the communication efficiency of various forms of oblivious transfer therefore represent an important class of open problems \cite{Ahlswede_Csiszar, Nascimento_Winter}. The capacity characterization of SPIR is a promising step in this  direction.

{\it Notation: For  $n_1, n_2\in\mathbb{Z}, n_1\leq n_2$, define the notation $[n_1:n_2]$ as the set $\{n_1,n_1+1,\cdots, n_2\}$.  For an index set $\mathcal{I}=\{i_1, i_2, \cdots, i_n\}$, with $i_1<i_2<\cdots<i_n$, the notation $A_{\mathcal{I}}$ represents the vector $(A_{i_1},A_{i_2},\cdots,A_{i_n})$. For an element $i_\theta$ in the set $\mathcal{I}=\{i_1, i_2, \cdots, i_n\}$, i.e., $i_\theta \in \mathcal{I}$, the notation $\overline{i_\theta}$ represents the complement of $\{i_\theta\}$, i.e., $\overline{i_\theta} \define \{i_1, \cdots, i_{\theta-1}, i_{\theta+1}, \cdots, i_n\}$. 
}

\section{Problem Statement}\label{sec:model}
Consider $K$ independent messages $W_1, \cdots, W_K, W_k \in \mathbb{F}_p^{l_k L \times 1}, k \in [1:K], l_k \in \mathbb{Z}_+, L \in \mathbb{Z}_+$, where $W_k$ is represented as an $l_k L \times 1$ vector comprised of $l_k L$ i.i.d. uniform symbols from a finite field $\mathbb{F}_p$ for a prime $p$. In $p$-ary units,
\begin{eqnarray}
&& H(W_1, \cdots, W_K) = H(W_1) + \cdots + H(W_K), \label{h1}\\
&& H(W_k) =  l_k L, \forall k \in [1:K]. 
\label{h2}
\end{eqnarray}
There are $N$ databases. Each database stores all the messages $W_1, \cdots, W_K$. 

\begin{figure}[h]
\begin{center}
  \includegraphics[scale = 0.61]{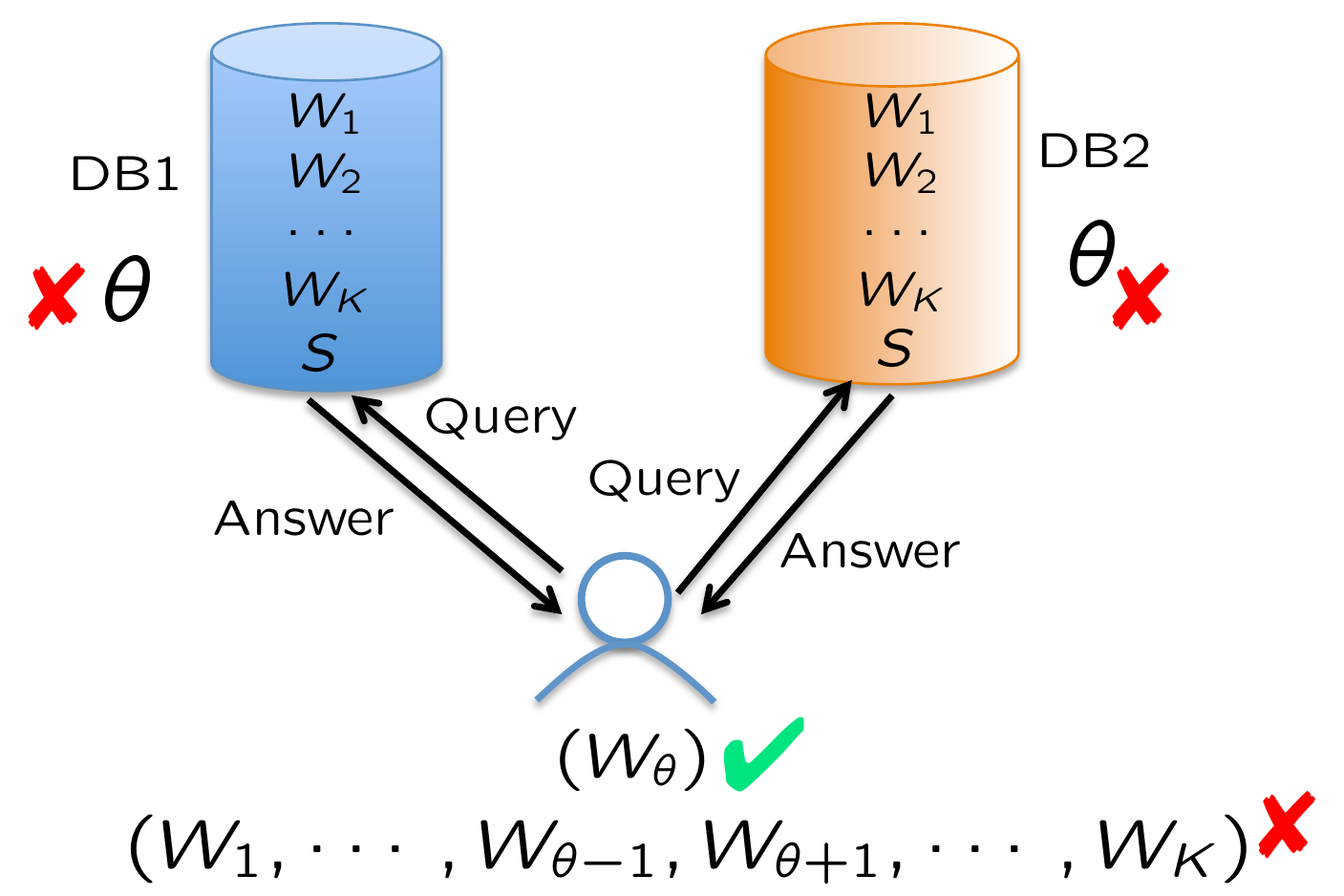}
  \caption{The SPIR problem with $K$ messages and 2 databases.}
  \label{spirs}
\end{center}
\end{figure}

Let us use $\mathcal{F}$ to denote a random variable privately generated by the user, whose realization is not available to the servers. $\mathcal{F}$ represents the randomness in the strategies followed by the user. 
The user privately generates $\theta$ uniformly from $[1:K]$ and wishes to retrieve $W_\theta$ privately. The databases do not want to give out any information beyond the one message of the user's choosing ($W_\theta$).  In order to achieve database-privacy, we assume that the databases share a common random variable $S$ that is not known to the user. It has been shown that without such common randomness, SPIR is not feasible \cite{SymPIR}. For a pictorial illustration of an example of the SPIR problem with $K$ messages and 2 databases, see Figure \ref{spirs}.
$\mathcal{F}$ 
is generated independently and before the realizations of the messages, the common randomness or the desired message index are known, so that
{\setlength{\mathindent}{0cm}
\begin{eqnarray}
H(\theta, \mathcal{F}, W_1, \cdots, W_K, S)  = H(\theta) + H(\mathcal{F}) + H(W_1) + \cdots + H(W_K) + H(S). \label{indep}
\end{eqnarray}
}

Suppose $\theta = k$. In order to retrieve message $W_k, k \in [1:K]$ privately, the user  privately generates $N$ queries $Q_1^{[k]}, \cdots, Q_N^{[k]}$. 
\begin{eqnarray}
H(Q_1^{[k]}, \cdots, Q_N^{[k]}|\mathcal{F}) = 0, \forall k \in [1:K]
\label{query_det}.
\end{eqnarray}
The user  sends query $Q_n^{[k]}$ to the $n$-th database, $n\in[1:N]$. Upon receiving $Q_n^{[k]}$, the $n$-th database generates an answering string $A_n^{[k]}$, which is a function of $Q_n^{[k]}$, all messages $W_1, \cdots, W_K$, and the common randomness $S$, 
\begin{eqnarray}
H(A_n^{[k]} | Q_n^{[k]}, W_1, \cdots, W_K, S) = 0.
\label{ansdet}
\end{eqnarray}
Each database returns to the user its answer $A_n^{[k]}$. 

From all the information that is now available to the user ($Q_{1:N}^{[k]}, A_{1:N}^{[k]}, \mathcal{F}$), the user decodes the desired message $W_k$ according to a decoding rule that is specified by the SPIR scheme. Let $P_e$ denote the probability of error achieved with the specified decoding rule.

To protect the user's privacy, the $K$ strategies must be indistinguishable (identically distributed) from the perspective of any individual database, 
i.e., the following user-privacy constraint must be satisfied\footnote{The User-Privacy constraint is equivalently expressed as $I(\theta; Q_{n}^{[\theta]}, A_{n}^{[\theta]}, W_{1:K}, S) = 0$.},
\begin{eqnarray}
\mbox{[User-Privacy]} ~~(Q_{n}^{[k]}, A_{n}^{[k]}, W_{1:K}, S) \sim (Q_{n}^{[k']}, A_{n}^{[k']}, W_{1:K}, S),\notag\\
\forall k, k' \in [1:K], \forall n \in [1:N].
\label{privacy}
\end{eqnarray}

Symmetric PIR also requires protecting the privacy of the database, i.e., it must be ensured that the user learns nothing more than the desired message $W_{k}$. So the vector  $W_{\overline{k}} = (W_1, \cdots, W_{k-1}, W_{k+1}, \cdots, W_K)$, must be independent of all the information available to the user. Thus, the following database-privacy constraint must be satisfied:
\begin{eqnarray}
\mbox{[DB-Privacy]} ~~ I(W_{\overline{k}}~; Q_{1:N}^{[k]}, A_{1:N}^{[k]}, \mathcal{F}) = 0, \forall k \in [1:K]. \label{noleak}
\end{eqnarray}

The SPIR rate of $W_k$ characterizes 
the amount of desired information retrieved per downloaded symbol, and is defined as follows.
\begin{eqnarray}
R_k \define \frac{l_k L}{D}.
\label{capacity}
\end{eqnarray}
where 
$D$ is the maximum value of the total number of symbols downloaded by the user from all the databases.

A rate tuple $(R_1, \cdots, R_K)$ is said to be $\epsilon$-error achievable if  $\forall k \in [1:K]$, there exists a sequence of PIR schemes, indexed by $L$, where the rate of $W_k$ is greater than or equal to $R_k$ 
and $P_e \rightarrow 0$ as $L \rightarrow \infty$. Note that for such a sequence of SPIR schemes, from Fano's inequality, we must have
\begin{eqnarray}
\mbox{[Correctness]} ~o(L) &=& \frac{1}{L} H(W_k | Q_{1:N}^{[k]}, A_{1:N}^{[k]}, \mathcal{F}) \\\
&\overset{(\ref{query_det})}{=}& \frac{1}{L} H(W_k | A_{1:N}^{[k]}, \mathcal{F}) \label{corr}
\end{eqnarray}
where $o(L)$ represents a term whose value approaches zero as $L$ approaches infinity.
The closure of the set of all $\epsilon$-error achievable rate tuples is called the capacity region $\mathcal{C}$.


\section{Results}\label{sec:main}
\subsection{Capacity of SPIR}\label{sec:spir}
In the typical setting of SPIR, the sizes of the messages are the same, i.e., $l_k = 1, \forall k \in [1:K]$ and the rate (refer to (\ref{capacity})) of each message is the same. 
Then the capacity region is characterized by one single parameter, i.e., the supremum of the achievable rate, named the capacity. We denote the capacity as $C$.

When there is only $K = 1$ message, note that the database-privacy constraint is satisfied trivially, so that SPIR reduces to the PIR setting and the capacity is $1$. For $K\geq 2$, it is known that some common randomness $S$ is necessary for the feasibility of SPIR. Let us define $\rho$ as the amount of common randomness relative to the message size
\begin{eqnarray}
\rho&=&\frac{H(S)}{H(W)}=\frac{H(S)}{L} \label{rho}
\end{eqnarray}
The capacity should depend on $\rho$, and because availability of common randomness  at the databases is a non-trivial requirement, this dependence is of some interest. 

When there is only $N = 1$ database, it is easy to see that the database-privacy constraint, the user-privacy constraint and correctness constraint conflict with each other such that SPIR is not feasible and the capacity is zero. The reason is as follows. First, because of the user-privacy constraint (\ref{privacy}), the answer from the only database $A_1^{[k]}$ is identically distributed for all $k \in [1:K]$. Second, from the correctness constraint (\ref{corr}), from $A_1^{[k]}, \mathcal{F}$, one can decode $W_k$. Combining these two facts, we have that from $A_1^{[k]}, \mathcal{F}$, one can decode all messages $W_1, \cdots, W_K$. This contradicts the database-privacy constraint (\ref{noleak}). Therefore, when $N = 1$ and $K \geq 2$, SPIR is not feasible.

The following theorem states 
the capacity of SPIR, when we have $N \geq 2$ databases and $K \geq 2$ messages.

\begin{theorem}\label{thm:download}
{\it For SPIR with $K \geq 2$ messages and $N \geq 2$ databases, the capacity is
\begin{eqnarray}
C_{\mbox{\footnotesize SPIR}} = 
\left\{
\begin{array}{c}
1 - 1/N ~~~\mbox{if} ~~\rho \geq \frac{1}{N-1} 
\\
0 ~~~~~~~~~\mbox{otherwise}
\end{array}
\right.
\end{eqnarray}}
\end{theorem}

The following observations place Theorem \ref{thm:download} in perspective.
\begin{enumerate}
\item 
We notice a surprising threshold phenomenon in the dependence of SPIR capacity, $C_{\mbox{\footnotesize SPIR}}$, on the amount of common randomness $\rho$. When $\rho < \frac{1}{N-1}$, SPIR is not feasible and $C_{\mbox{\footnotesize SPIR}}=0$. However, when $\rho\geq\frac{1}{N-1}$, SPIR is not only possible, but the rate can immediately be increased to the maximum possible, i.e., the capacity. Therefore, the minimum  common randomness required to achieve any positive rate is already sufficient to achieve the capacity of SPIR. A pictorial illustration of the SPIR capacity and its dependency on the amount of common randomness appears in Figure \ref{cspir}.

\begin{figure}[h]
\begin{center}
  \includegraphics[scale = 0.55]{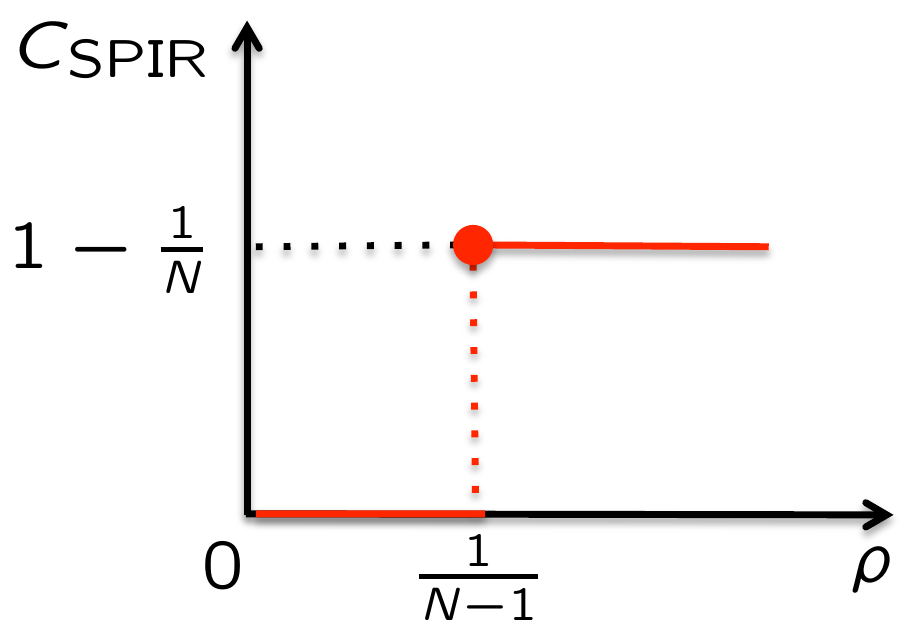}
  \caption{SPIR capacity.}
  \label{cspir}
\end{center}
\end{figure}

\item
The capacity of SPIR is independent of the number of messages, $K$. 

\item
When the capacity is non-zero, the capacity is strictly increasing in the number of databases, $N$, and when $N$ approaches infinity, the capacity approaches 1. 

\item
It is interesting to compare the capacity of SPIR and the capacity of PIR \cite{Sun_Jafar_PIR},
\begin{eqnarray}
C_{\footnotesize \mbox{PIR}} = \left(1 +1/N + 1/{N^2} + \cdots +1/{N^{K-1}}\right)^{-1}.
\end{eqnarray} 
We see that the capacity of SPIR is strictly smaller than the capacity of PIR (the additional requirement of preserving database-privacy strictly hurts) and the capacity of PIR approaches the capacity of SPIR when the number of messages, $K$, approaches infinity (in the large number of messages regime, the penalty vanishes), i.e., $C_{\footnotesize \mbox{PIR}} > C_{\mbox{\footnotesize SPIR}}$ for any finite $K$ and $C_{\footnotesize \mbox{PIR}} \rightarrow C_{\mbox{\footnotesize SPIR}}$ when $K \rightarrow \infty$. 

\item 
The achievable scheme presented in Section \ref{sec:ach} has exactly zero error. Further, In the achievability proof for Theorem \ref{thm:download}, the message size is $N-1$ bits per message. Therefore, to achieve capacity, message size is not required to approach infinity. By employing the scheme multiple times, we know that when message size is equal to an integer multiple of $N-1$ bits, the capacity is achieved as well. When the message size is not equal to an integer multiple of $N-1$ bits, it turns out that there is a penalty in the form of a ceiling operation. This extension of SPIR to finite length messages is considered in Theorem \ref{thm:finite}, to be presented in Section \ref{sec:finite}. 

\item We note that the converse (upper bound, presented in Section \ref{sec:converse}) holds for arbitrary message size $L$  when we require exactly zero error, by replacing the $o(L)$ terms  with zero.

\item The extension to unequal message sizes is considered in Section \ref{sec:region}. 
\end{enumerate}

In the following sections, we relax each one of the two assumptions by itself, i.e., equal message sizes and message length $L$ going to infinity. 

\subsection{Capacity Region of SPIR}\label{sec:region}
In this section, we relax the assumption of equal message sizes, i.e., $l_k, \forall k \in [1:K]$ are arbitrary. Therefore, going beyond the (symmetric) capacity, we wish to characterize the capacity region of SPIR. 

When we only have $K=1$ message, similar to the previous section, the capacity region is characterized by the capacity of one message, which is 1. When we only have $N = 1$ database and $K \geq 2$ messages, similar to the previous section, SPIR is not feasible and the capacity region is the zero vector. 
Therefore, we consider $K \geq 2$ messages and $N \geq 2$ databases, where the capacity region of SPIR is characterized in the following theorem. Here the amount of common randomness is normalized with respective to the largest message size.
\begin{eqnarray}
\rho&=&\frac{H(S)}{\max_{i: i \in [1:K]} H(W_i)}=\frac{H(S)}{\max_{i:i\in[1:K]} l_i L}, 
\end{eqnarray}

\begin{theorem}\label{thm:region}
{\it For SPIR with $K \geq 2$ messages and $N \geq 2$ databases, the capacity region $\mathcal{C}$ is
\begin{eqnarray}
\mathcal{C} = \left\{(R_1, \cdots, R_K) :
R_k \leq  
\frac{l_k}{\max_i l_i}(1 - \frac{1}{N}),
\forall k \in [1:K]
\right\}, \mbox{if} ~~\rho \geq \frac{1}{N-1}
\end{eqnarray}}
and the zero vector 
otherwise.
\end{theorem}

{\it Remark: The optimal (minimum) normalized download cost $D/L  = l_k / R_k = \max_i l_i \frac{N}{N-1}$ is the same for each message.}

\subsection{Capacity of Finite Length SPIR}\label{sec:finite}
In this section, we again assume that all messages have the same length, but relax the assumption that $L$ approaches infinity. Instead, we assume that  $L$ is an arbitrary finite value. As $L$ is finite, we consider zero error achievable rates and define its supremum as zero error capacity, denoted as $C_o$. This setting can be obtained from the general problem statement by setting $l_k = 1, \forall k \in [1:K]$, and $L$ finite.

Similar to Section \ref{sec:spir}, we restrict to $K \geq 2$ and $N \geq 2$ cases as the problem is trivial when $K=1$ or $N = 1$.
The capacity of finite length SPIR is characterized in the following theorem. The relative size of the common randomness, $\rho$, is defined as in (\ref{rho}).

\begin{theorem}\label{thm:finite}
{\it For SPIR with $K \geq 2$ messages, $N \geq 2$ databases, where each message is of size $L \in \mathbb{Z}_+$ symbols, the zero error capacity is
\begin{eqnarray}
C_{o,\mbox{\footnotesize LSPIR}} = 
\left\{
\begin{array}{cc}
L/\lceil\frac{L}{C_{\mbox{\tiny SPIR}}}\rceil = L/\lceil\frac{L}{1 - 1/N}\rceil  & \mbox{if} ~\rho \geq \frac{\lceil L/(N-1) \rceil}{L}
\\
0 & \mbox{otherwise}
\end{array}
\right.
\end{eqnarray}}
\end{theorem}

%
%

\section{Proofs}
\subsection{Proof of Theorem \ref{thm:download}}
 
\subsubsection{Achievability}\label{sec:ach}
In this section, we present the scheme that achieves rate $1 - 1/N$,  when $\rho = 1/(N-1)$. To this end, we assume each message consists of $N-1$ bits and each answering string is 1 bit. Specifically, we assume $W_k = (x_{k, 1}, \cdots, x_{k, N-1}),\forall k \in [1:K]$ where each $x_{k, i}, i \in [1:N-1]$ is one bit. We further assume the entropy of the common random variable $S$ is $1$ bit, i.e., $S$ is uniformly distributed over $\{0,1\}$. Note that $S$ is independent of the messages.

Next we specify the queries. To retrieve $W_{k}$ privately, the user first generates a random  vector of length $(N-1)K$, $[h_{1,1}, \cdots, h_{1,N-1}, \cdots, h_{k,1}, \cdots, h_{K, N-1}]$, where each element is uniformly distributed over $\{0,1\}$. Then the queries are set as follows.
\begin{eqnarray}
Q_1^{[k]} &=& [h_{1,1}, \cdots, h_{k,1}, \cdots, h_{k,N-1}, \cdots, h_{K, N-1}] \notag\\
Q_2^{[k]} &=& [h_{1,1}, \cdots, h_{k,1}+1, \cdots, h_{k,N-1}, \cdots, h_{K, N-1}] \notag\\
&\cdots&\notag\\
Q_N^{[k]} &=& [h_{1,1}, \cdots, h_{k,1}, \cdots, h_{k,N-1} + 1, \cdots, h_{K, N-1}] \label{eq:q}
\end{eqnarray}
The answering strings are generated by using the query vector as the combining coefficients and producing the corresponding linear combination of message bits. We further add the common random variable to each answer.
\begin{eqnarray}
A_1^{[k]} &=& \sum_{j=1}^K \sum_{i=1}^{N-1} h_{j,i} x_{j,i} + S\notag\\
A_2^{[k]} &=& \sum_{j=1}^K \sum_{i=1}^{N-1} h_{j,i} x_{j,i}  + x_{k,1} + S\notag\\
&\cdots&\notag\\
A_N^{[k]} &=& \sum_{j=1}^K \sum_{i=1}^{N-1} h_{j,i} x_{j,i} + x_{k,N-1} + S \label{eq:a}
\end{eqnarray}
The user obtains $ x_{k,i}, i \in [1:N-1]$ by subtracting $A_1^{[k]}$ from $A_{i+1}^{[k]}$. Therefore, the correctness condition is satisfied. 

Privacy of the user is guaranteed because each query is independent of the desired message index $k$. This is because regardless of the desired message index $k$, each of the query vectors $Q_n^{[k]}, \forall n$ is individually comprised of elements that are i.i.d. uniform over $\{0,1\}$. Thus, each database learns nothing about which message is requested. 

We now show that database-privacy is preserved as well.
\begin{eqnarray}
&& I(W_{\overline{k}}~; A_1^{[k]}, A_2^{[k]}, \cdots, A_N^{[k]}, Q_{1:N}^{[k]}, \mathcal{F}) \\
&=& I(W_{\overline{k}}~; A_1^{[k]}, A_1^{[k]} + x_{k,1}, \cdots, A_1^{[k]} + x_{k, N-1}, Q_{1:N}^{[k]}, \mathcal{F}) \\
&=& I(W_{\overline{k}}~; A_1^{[k]}, x_{k,1}, \cdots, x_{k, N-1}, Q_{1:N}^{[k]}, \mathcal{F}) \\
&=& I(W_{\overline{k}}~; A_1^{[k]}, W_{k}, Q_{1:N}^{[k]}, \mathcal{F}) \\
&\overset{(\ref{indep})(\ref{query_det})}{=}& I(W_{\overline{k}}~;  A_1^{[k]}| W_{k} , Q_{1:N}^{[k]}, \mathcal{F}) \\
&=& 0
\end{eqnarray}
where in each step, the transformation on the variables is invertible such that mutual information remains the same. The last step follows from the independence of the messages and the common randomness (refer to (\ref{indep})).

Note that because each answering string is $1$ bit and the message is $L=N-1$ bits, the rate achieved is $(N-1)/N=1-1/N$ which matches the capacity. Also note that only the minimum threshold amount of common randomness is utilized, i.e., $\rho=1/(N-1)$.\hfill\QED

\subsubsection{Converse}\label{sec:converse}
Although Theorem \ref{thm:download} restricts to the setting where $l_k = 1, \forall k \in [1:K]$, we do not assume this at the beginning in the proof of converse. This will make the converse general such that some of the intermediate steps can be used in the converse proofs of Theorem \ref{thm:region} and Theorem \ref{thm:finite} as well.

For the converse we allow any feasible SPIR scheme, and prove that its rate cannot be larger than $C_{\mbox{\footnotesize SPIR}}$. Let us start with two lemmas that will be used later in the proof.

\begin{lemma}\label{lemma:claim}
\begin{eqnarray}
H(A_n^{[k]}|W_{k},{Q}_n^{[k]})&=&H(A_n^{[k']}|W_{k},{Q}_n^{[k']}) \label{c1}\\
H(A_n^{[k]}|{Q}_n^{[k]})&=&H(A_n^{[k']}|{Q}_n^{[k']}) \label{c2}, ~~~ \forall  n\in [1:N]
\end{eqnarray}
\end{lemma}
\proof Since the proofs of (\ref{c1}) and (\ref{c2}) follow from the same arguments, here we will present only the proof of (\ref{c1}). From the User-Privacy constraint (\ref{privacy}) we know that $\forall k\in[1:K], \forall n \in [1:N]$, $I(\theta; A_n^{[\theta]},W_{k}, Q_n^{[\theta]})=0$. Therefore, we must have $\forall k'\in[1:K],$
\begin{eqnarray}
H(A_n^{[k]},W_{k},Q_n^{[k]})&=&H(A_n^{[k']},W_{k},Q_n^{[k']})\label{eq:l1}\\
H(W_{k},Q_n^{[k]})&=&H(W_{k}, Q_n^{[k']}) \label{eq:l2}
\end{eqnarray}
Combining (\ref{eq:l1}) and (\ref{eq:l2}), we obtain $H(A_n^{[k]}|W_{k},Q_n^{[k]})=H(A_n^{[k']}|W_{k},Q_n^{[k']})$.
\hfill\QED

\begin{lemma}\label{lemma:addq}
\begin{align}
H(A_n^{[k]} | W_{k}, \mathcal{F}, Q_n^{[k]}) = H(A_n^{[k]} | W_{k}, Q_n^{[k]}) \label{q1}, ~~~\forall n \in [1:N]
\end{align}
\end{lemma}
\proof Since 
\begin{align} 
&H(A_n^{[k]} | W_{k}, Q_n^{[k]}) - H(A_n^{[k]} | W_{k}, \mathcal{F}, Q_n^{[k]}) 
=
I(A_n^{[k]}; \mathcal{F} | W_{k}, Q_n^{[k]}) \geq 0,
\end{align} 
we only need to prove $I(A_n^{[k]}; \mathcal{F} | W_{k}, Q_n^{[k]}) \leq 0$. 
\begin{eqnarray} 
&& I(A_n^{[k]}; \mathcal{F} | W_{k}, Q_n^{[k]}) \\
&\leq& I(A_n^{[k]}, W_1, \cdots, W_K, S; \mathcal{F} | W_{k}, Q_n^{[k]}) \\
&=& I(W_1, \cdots, W_K, S; \mathcal{F} | W_{k}, Q_n^{[k]}) 
+ \underbrace{I(A_n^{[k]};\mathcal{F}| W_1, \cdots, W_K, S, W_{k}, Q_n^{[k]})}_{= 0} \label{q2} \\
&\leq& I(W_1, \cdots, W_K, S; \mathcal{F}, Q_n^{[k]}) \\
&=& 0 \label{q3}
\end{eqnarray}
where the second term in (\ref{q2}) is zero because of (\ref{ansdet}) and (\ref{q3}) follows from (\ref{indep}), (\ref{query_det}).
\hfill\QED

\paragraph{The proof for $R\leq C_{\mbox{\footnotesize SPIR}}$.}

{\color{black}
For every feasible SPIR scheme, we must satisfy the database-privacy constraint (\ref{noleak}), 
\begin{align}
0&=I(W_{\overline{k'}}~; A_1^{[k']}, \cdots, A_N^{[k']}, Q_1^{[k']}, \cdots, Q_N^{[k']}, \mathcal{F}) \\
\intertext{such that $\forall n \in [1:N], \forall k\in[1:K], k\neq k'$,}
0&= I(W_{k}; A_n^{[k']},{Q}_n^{[k']}) \\
&= H(A_n^{[k']}|{Q}_n^{[k']}) - H(A_n^{[k']}|W_{k},{Q}_n^{[k']}) \label{eq:same1}\\
&\overset{(\ref{c1})}{=} H(A_n^{[k']}|{Q}_n^{[k']}) - H(A_n^{[k]}|W_{k},{Q}_n^{[k]}) \label{eq:same}
\end{align}
Now, consider  the answering strings $A_1^{[k]}, \cdots, A_N^{[k]}$, from which we can decode $W_{k}$. 
\begin{align}
l_kL &= H(W_{k}) \overset{(\ref{indep})}{=} H(W_{k}|\mathcal{F})\\
&\overset{(\ref{corr})}{=} I(W_{k}; A_1^{[k]}, \cdots, A_N^{[k]}|\mathcal{F}) + o(L)L\\
&= H(A_1^{[k]}, \cdots, A_N^{[k]}|\mathcal{F}) - H(A_1^{[k]}, \cdots, A_N^{[k]}|W_{k}, \mathcal{F}) + o(L)L \\
&\overset{(\ref{query_det})}{\leq} H(A_1^{[k]}, \cdots, A_N^{[k]}|\mathcal{F}) - H(A_n^{[k]} | W_{k}, \mathcal{F}, Q_n^{[k]}) + o(L)L \\
&\overset{(\ref{q1})}{=} H(A_1^{[k]}, \cdots, A_N^{[k]}|\mathcal{F}) - H(A_n^{[k]} | W_{k}, Q_n^{[k]}) + o(L)L 
\\
&\overset{(\ref{eq:same})}{=} H(A_1^{[k]}, \cdots, A_N^{[k]}|\mathcal{F}) - H(A_n^{[k']} | {Q}_n^{[k']}) + o(L)L\\
&\overset{(\ref{c2})}{=} H(A_1^{[k]}, \cdots, A_N^{[k]}|\mathcal{F}) - H(A_n^{[k]} | {Q}_n^{[k]}) + o(L)L\\
&\overset{(\ref{query_det})}{\leq} H(A_1^{[k]}, \cdots, A_N^{[k]}|\mathcal{F}) - H(A_n^{[k]} | \mathcal{F}) + o(L)L\label{eq:last}
\end{align}
}

Adding (\ref{eq:last}) for all $n \in [1:N]$, we have
\begin{eqnarray}
N l_k L &\leq& N H(A_1^{[k]}, \cdots, A_N^{[k]}|\mathcal{F}) - \sum_{n \in [1:N]} H(A_n^{[k]} | \mathcal{F}) + o(L)L \\
&\leq&  \left(N-1\right)H(A_1^{[k]}, \cdots, A_N^{[k]}|\mathcal{F})  \label{eq:ee} + o(L)L 
\\
&\leq&  \left(N-1 \right) \sum_{n=1}^N H(A_n^{[k]}) + o(L)L\\
&\leq& \left(N-1 \right) D + o(L)L
\label{eq:asame} \\
R_k &=& \frac{l_k L}{D
} \leq 1 - \frac{1}{N} ~~~{\mbox{(Letting $L\rightarrow \infty$)}}
\end{eqnarray}
Thus, the rate of any feasible SPIR scheme cannot be more than $C_{\mbox{\footnotesize SPIR}}$. 


\paragraph{The proof for $\rho \geq 1/(N-1)$.}

Suppose a feasible SPIR scheme exists that achieves a non-zero SPIR rate. Then we will show that it must have $\rho \geq 1/(N-1)$. Consider the answering strings $A_1^{[k]}, \cdots, A_N^{[k]}$, from which we can decode $W_{k}$. From the database-privacy constraint, we have
\begin{eqnarray}
0 &=& 
I(W_{\overline{k}}~; A_1^{[k]}, \cdots, A_N^{[k]}, \mathcal{F}) \\
&\overset{(\ref{indep})}{=}& I(W_{\overline{k}}~; A_1^{[k]}, \cdots, A_N^{[k]} | \mathcal{F}) \\
&\overset{(\ref{corr})}{=}& I(W_{\overline{k}}~; A_1^{[k]}, \cdots, A_N^{[k]}, W_{k} | \mathcal{F}) + o(L)L\label{eq:e1}\\
&\overset{(\ref{indep})}{=}& I(W_{\overline{k}}~; A_1^{[k]}, \cdots, A_N^{[k]} | W_{k}, \mathcal{F}) + o(L)L\label{eq:e2} \\
&\geq& I(W_{\overline{k}}~; A_n^{[k]} | W_{k}, \mathcal{F}) + o(L)L\\
&=& H(A_n^{[k]}|W_{k}, \mathcal{F}) - H(A_n^{[k]}|W_{1}, \cdots, W_K, \mathcal{F}) + o(L)L\\
&\overset{(\ref{query_det})(\ref{ansdet})}{=}& H(A_n^{[k]}|W_{k}, \mathcal{F}) - H(A_n^{[k]}|W_{1}, \cdots, W_K, \mathcal{F}) \notag \\
&&~+ H(A_n^{[k]}|W_{1}, \cdots, W_K, \mathcal{F}, S) \label{eq:e3}  + o(L)L\\
&=& H(A_n^{[k]}|W_{k}, \mathcal{F}) - I(S;A_n^{[k]}|W_{1}, \cdots, W_K, \mathcal{F}) + o(L)L\\
&\overset{(\ref{query_det})}{\geq}& H(A_n^{[k]}|W_{k}, \mathcal{F}, Q_n^{[k]}) - H(S) + o(L)L\\
&\overset{(\ref{q1})}{=}& H(A_n^{[k]}|W_{k}, Q_n^{[k]}) - H(S) \label{eq:e4} + o(L)L\\
&\overset{(\ref{eq:same})}{=}& H(A_n^{[k']}| Q_n^{[k']}) - H(S) \label{eq:e5} + o(L)L\\
&\overset{(\ref{c2})}{=}& H(A_n^{[k]}| Q_n^{[k]}) - H(S) + o(L)L\label{eq:e6}
\end{eqnarray}

Adding (\ref{eq:e6}) for $n \in [1:N]$, we have
\begin{eqnarray}
0 &\geq& \sum_{n \in [1:N]} H(A_{n}^{[k]}|{Q}_n^{[k]}) - N H(S) + o(L)L\\
&\geq& H(A_1^{[k]}, \cdots, A_N^{[k]}|\mathcal{F}) - N H(S) + o(L)L\\
&\overset{(\ref{eq:ee})}{\geq}& \frac{N}{N-1} l_k L - N H(S) \label{eq:e7} + o(L)L\\
\Rightarrow H(S) &\geq& \frac{1}{N-1} l_k L + o(L)L\\
\Rightarrow \rho &=& \frac{H(S)}{l_k L} \geq \frac{1}{N-1} \label{eq:rhobound} ~~~\mbox{(Letting $L \rightarrow \infty$)}
\end{eqnarray} 
Thus, the amount of common randomness relative to the message size of any feasible SPIR scheme cannot be less than $1/(N-1)$.

\subsection{Proof for Theorem \ref{thm:region}}
\subsubsection{Achievability}\label{sec:proof_region_ach}
Without loss of generality, we assume that $l_1 \leq l_2 \leq \cdots \leq l_K$. As a result, we need to prove that for $W_k$, rate $(1 - 1/N)l_k/l_K$ is achievable,  when $\rho = 1/(N-1)$. Here is such a scheme. We set $L = N-1$. We divide each message into $K$ sub-messages and for each sub-message, we use the SPIR scheme (\ref{eq:q}), (\ref{eq:a}). That is
\begin{eqnarray}
W_k &=& (W_k(1), W_k(2), \cdots, W_k(K)) \\
W_k(i) &\in& \mathbb{F}_2^{(l_i - l_{i-1})L \times 1} ~~(\mbox{Define $l_0 = 0$})
\end{eqnarray}
Note that $H(W_k) = l_k L$, so we set $W_k(i)$ to zero vectors, 
when $i \in [k+1:K]$. For sub-messages $(W_1(i), W_2(i), \cdots, W_K(i))$, we employ the SPIR scheme (\ref{eq:q}), (\ref{eq:a}) $l_k - l_{k-1}$ times independently. Therefore the number of bits downloaded, denoted as $D_k(i)$, and the amount of common randomness used, $H(S(i)$), are obtained as follows. $\forall i \in [1:K],$
\begin{eqnarray}
D_k(i) &=& (l_k - l_{k-1}) L/(1- 1/N) = (l_k - l_{k-1})N \label{eq:dk}\\
H(S(i)) &=& (l_k - l_{k-1}) L/(N-1) = l_k - l_{k-1} \label{eq:sk}
\end{eqnarray}
The schemes for each sub-message are also independent. As our scheme is a concatenation of multiple independent correct and private SPIR schemes, the overall concatenated scheme is also correct and private. The proof is similar to Theorem 4 of \cite{Sun_Jafar_LPIR} and is thus omitted.

Finally, the rate and the amount of common randomness are as follows.
\begin{eqnarray}
R_k &=& \frac{H(W_k)}{D} =  \frac{\sum_{i=1}^K H(W_k(i))}{\sum_{i=1}^K D_k(i)} \overset{(\ref{eq:dk})}{=} \frac{l_k L}{l_K N} = \frac{l_k }{l_K } \left(1- \frac{1}{N}\right)\\
\rho &=& \frac{H(S)}{H(W_K)} = \frac{\sum_{i=1}^K H(S(i))}{l_K L} \overset{(\ref{eq:sk})}{=} \frac{l_K}{l_k L} = \frac{1}{N-1}
\end{eqnarray}
Therefore, the achievability of Theorem \ref{thm:region} is proved.


\subsubsection{Converse}\label{sec:proof_region_converse}
The converse proof is almost identical to that of Theorem \ref{thm:download}. Note that (\ref{eq:asame}) holds for all $l_k$ and all $k$, we have 
\begin{eqnarray}
N \max_i l_i  &\leq& (N-1) D/L  \\
\Rightarrow R_k &=& l_kL /D \leq \frac{l_k }{\max_i l_i } \left(1- \frac{1}{N}\right)
\end{eqnarray}
Therefore the rate bound is proved. The common randomness bound is identical to (\ref{eq:rhobound}). Note that (\ref{eq:rhobound}) holds for all $k \in [1:K]$. 

\subsection{Proof for Theorem \ref{thm:finite}}
\subsubsection{Achievability}\label{sec:proof_finite_ach}
Suppose $L = G_1 (N-1) + L_1$, where $G_1 = \lfloor L/(N-1)\rfloor$ and $L_1 \in [0: N-2]$. Note that the capacity achieving scheme for SPIR when $L$ is not restricted is based on dividing the messages to blocks of length $N-1$ (refer to Theorem \ref{thm:download}). The optimal scheme for finite $L$ setting is constructed by first using the capacity achieving SPIR scheme $G_1$  times to retrieve $G_1(N-1)$ bits, and then for the remaining $L_1$ bits, we use the capacity achieving SPIR schemes with only $L_1 + 1 \leq N$ databases (say, the first $L_1 + 1$ databases), if $L_1 \geq 1$. Otherwise if $L_1 = 0$, then we are done. Note that for the SPIR scheme that uses only $L_1 + 1$ databases, the rate is $1 - 1/(L_1 +1)$, the message size is $L_1 + 1 - 1 = L_1$ bits, and the common randomness ratio is $\rho = 1/(L_1 + 1 -1) = 1/L_1$. Therefore, overall, the rate and the amount of common randomness are as follows.
\begin{eqnarray}
R &=& \left\{ \begin{array}{cc}  \frac{G_1(N-1)}{ G_1 N} = 1-\frac{1}{N}, & \mbox{if}~ L_1 = 0, \\
\frac{G_1(N-1) + L_1}{ G_1 N + L_1 + 1}, & \mbox{otherwise.}
\end{array} \right. \label{eq:finite_r} \\ 
\rho &=& \left\{ \begin{array}{cc}  \frac{G_1}{ G_1 (N-1)} = \frac{1}{N-1}, & \mbox{if}~ L_1 = 0, \\
\frac{G_1 + 1}{G_1 (N-1) + L_1} = \frac{\lfloor L/(N-1) \rfloor+ 1}{L},
 & \mbox{otherwise.}
\end{array} \right. \\
&=& \frac{\lceil L/(N-1) \rceil}{L}
\end{eqnarray}
Next, we prove that the rate achieved in (\ref{eq:finite_r}) matches that in Theorem \ref{thm:finite}, i.e., $L/\lceil\frac{L}{1 - 1/N}\rceil$. When $L_1 = 0$ ($L$ is an integer multiple of $N-1$), the claim follows trivially.
Hereafter, we consider $L_1 > 0$. It suffices to show that the download cost, $D = L/R = G_1N + L_1 + 1$, satisfies $D \in [\frac{L}{1 - 1/N}, \frac{L}{1 - 1/N} +1)$. In the converse of Theorem \ref{thm:download}, we have showed that for arbitrary $L$ and all SPIR schemes, $D \geq \frac{L}{1 - 1/N}$ holds. So we are left to show that $D < \frac{L}{1 - 1/N} +1$. 
\begin{eqnarray}
D &=& G_1N + L_1 + 1 \\
&<& \frac{G_1 (N-1) + L_1}{1- 1/N} +1 ~~(N \geq 2)\\
&=& \frac{L}{1 - 1/N} +1
\end{eqnarray}
Therefore the achievability proof is complete.

\subsubsection{Converse}\label{sec:proof_finite_converse}
We show for fixed finite $L$, the achievable rate $R \leq L/\lceil\frac{L}{1 - 1/N}\rceil$. 
Equivalently, it suffices to prove that the download cost $D = L/R \geq \lceil\frac{L}{1 - 1/N}\rceil \geq \frac{L}{1 - 1/N}$, which follows from Theorem \ref{thm:download}. Note that Theorem \ref{thm:download} holds when we require exactly zero error. Also note that since the downloads are assumed to be in terms of symbols from the same field as the message symbols, the download cost must be an integer value.

We are left to prove the common randomness bound. Similar to the download cost, which is restricted to be integers, in the finite length regime the amount of common randomness, $\rho L$ is restricted to take integer values as well. Therefore, from (\ref{eq:rhobound}), we have $\rho L \geq {\lceil L/(N-1) \rceil}$ (for this setting, $l_k = 1$).

\section{Conclusion}\label{sec:conc}
For $K$ messages and $N$ databases, the capacity of SPIR was shown to be $C=1-1/N$. In order to achieve any positive rate for SPIR, the minimum amount of common randomness needed among the databases was shown to be $1/(N-1)$ bits per message bit. Remarkably, this is also sufficient to achieve the capacity of SPIR. The insights extend to settings with unequal message sizes and finite length messages. 



\bibliographystyle{IEEEtran}
\bibliography{Thesis}

\begin{thebibliography}{10}
\providecommand{\url}[1]{#1}
\csname url@samestyle\endcsname
\providecommand{\newblock}{\relax}
\providecommand{\bibinfo}[2]{#2}
\providecommand{\BIBentrySTDinterwordspacing}{\spaceskip=0pt\relax}
\providecommand{\BIBentryALTinterwordstretchfactor}{4}
\providecommand{\BIBentryALTinterwordspacing}{\spaceskip=\fontdimen2\font plus
\BIBentryALTinterwordstretchfactor\fontdimen3\font minus
  \fontdimen4\font\relax}
\providecommand{\BIBforeignlanguage}[2]{{%
\expandafter\ifx\csname l@#1\endcsname\relax
\typeout{** WARNING: IEEEtran.bst: No hyphenation pattern has been}%
\typeout{** loaded for the language `#1'. Using the pattern for}%
\typeout{** the default language instead.}%
\else
\language=\csname l@#1\endcsname
\fi
#2}}
\providecommand{\BIBdecl}{\relax}
\BIBdecl

\bibitem{PIRfirst}
B.~Chor, O.~Goldreich, E.~Kushilevitz, and M.~Sudan, ``Private information
  retrieval,'' in \emph{Proceedings of the 36th Annual Symposium on Foundations
  of Computer Science}, 1995, pp. 41--50.

\bibitem{PIRfirstjournal}
B.~Chor, E.~Kushilevitz, O.~Goldreich, and M.~Sudan, ``{Private Information
  Retrieval},'' \emph{Journal of the ACM (JACM)}, vol.~45, no.~6, pp. 965--981,
  1998.

\bibitem{Sun_Jafar_PIR}
H.~Sun and S.~A. Jafar, ``{The Capacity of Private Information Retrieval},''
  \emph{arXiv preprint arXiv:1602.09134}, 2016.

\bibitem{SymPIR}
Y.~Gertner, Y.~Ishai, E.~Kushilevitz, and T.~Malkin, ``Protecting data privacy
  in private information retrieval schemes,'' in \emph{Proceedings of the
  thirtieth annual ACM symposium on Theory of computing}.\hskip 1em plus 0.5em
  minus 0.4em\relax ACM, 1998, pp. 151--160.

\bibitem{William}
W.~Gasarch, ``{A Survey on Private Information Retrieval},'' in \emph{Bulletin
  of the EATCS}, 2004.

\bibitem{Yekhanin}
S.~Yekhanin, ``{Private Information Retrieval},'' \emph{Communications of the
  ACM}, vol.~53, no.~4, pp. 68--73, 2010.

\bibitem{LDC}
J.~Katz and L.~Trevisan, ``On the efficiency of local decoding procedures for
  error-correcting codes,'' in \emph{Proceedings of the thirty-second annual
  ACM symposium on Theory of computing}.\hskip 1em plus 0.5em minus 0.4em\relax
  ACM, 2000, pp. 80--86.

\bibitem{YekhaninPhd}
S.~Yekhanin, ``{Locally Decodable Codes and Private Information Retrieval
  Schemes},'' Ph.D. dissertation, Massachusetts Institute of Technology, 2007.

\bibitem{Batch}
Y.~Ishai, E.~Kushilevitz, R.~Ostrovsky, and A.~Sahai, ``{Batch codes and their
  applications},'' in \emph{Proceedings of the thirty-sixth annual ACM
  symposium on Theory of computing}.\hskip 1em plus 0.5em minus 0.4em\relax
  ACM, 2004, pp. 262--271.

\bibitem{Ishai_Kushilevitz}
Y.~Ishai and E.~Kushilevitz, ``{On the hardness of information-theoretic
  multiparty computation},'' in \emph{Advances in Cryptology-EUROCRYPT
  2004}.\hskip 1em plus 0.5em minus 0.4em\relax Springer, 2004, pp. 439--455.

\bibitem{Rabin}
M.~O. Rabin, ``How to exchange secrets with oblivious transfer.'' 1981.

\bibitem{Even_OT}
S.~Even, O.~Goldreich, and A.~Lempel, ``A randomized protocol for signing
  contracts,'' \emph{Communications of the ACM}, vol.~28, no.~6, pp. 637--647,
  1985.

\bibitem{Killian}
J.~Kilian, ``Founding crytpography on oblivious transfer,'' in
  \emph{Proceedings of the twentieth annual ACM symposium on Theory of
  computing}.\hskip 1em plus 0.5em minus 0.4em\relax ACM, 1988, pp. 20--31.

\bibitem{Ishai_Prabhakaran_Sahai}
Y.~Ishai, M.~Prabhakaran, and A.~Sahai, ``Founding cryptography on oblivious
  transfer--efficiently,'' in \emph{Annual International Cryptology
  Conference}.\hskip 1em plus 0.5em minus 0.4em\relax Springer, 2008, pp.
  572--591.

\bibitem{Ahlswede_Csiszar}
R.~Ahlswede and I.~Csisz{\'a}r, ``On oblivious transfer capacity,'' in
  \emph{Information Theory, Combinatorics, and Search Theory}.\hskip 1em plus
  0.5em minus 0.4em\relax Springer, 2013, pp. 145--166.

\bibitem{Nascimento_Winter}
A.~C. Nascimento and A.~Winter, ``On the oblivious-transfer capacity of noisy
  resources,'' \emph{IEEE Transactions on Information Theory}, vol.~54, no.~6,
  pp. 2572--2581, 2008.

\bibitem{Sun_Jafar_LPIR}
H.~Sun and S.~A. Jafar, ``{Optimal Download Cost of Private Information
  Retrieval for Arbitrary Message Length},'' \emph{arXiv preprint
  arXiv:1610.03048}, 2016.

\end{thebibliography}
\end{document}